\documentclass[twocolumn]{aastex62}

\accepted{June 11, 2018}
\usepackage{amsmath}

\submitjournal{The Astrophysical Journal}

\shorttitle{The near-IR spectrum of the Serpens S68N Class 0 protostar}
\shortauthors{Greene, Gully-Santiago, \& Barsony}

\begin{document}

\title{Detection of Photospheric Features in the Near-Infrared Spectrum of a Class 0 Protostar }

\correspondingauthor{Thomas Greene}
\email{tom.greene@nasa.gov}

\author[0000-0002-8963-8056]{Thomas P. Greene}
\affiliation{NASA Ames Research Center\\
 Space Science and Astrobiology Division\\ 
 M.S. 245-6\\
 Moffett Field, CA 94035, USA\\}

\author[0000-0002-4020-3457]{Michael A. Gully-Santiago}
\affiliation{NASA Ames Research Center and Bay Area Environmental Research Institute\\
 Moffett Field, CA 94035, USA\\}

\author{Mary Barsony}
\affiliation{SETI Institute\\
189 Bernardo Ave., 2nd Floor\\
Mountain View, CA 94043, USA}

\begin{abstract}

We present a near-infrared $K$-band $R \simeq 1500$ Keck spectrum of S68N, a Class 0 protostar in the Serpens molecular cloud. The spectrum shows a very red continuum, CO absorption bands, weak or non-existent atomic metal absorptions, and H$_2$ emission lines. The near-IR H$_2$ emission is consistent with excitation in shocks or by X-rays but not by UV radiation. We model the absorption component as a stellar photosphere plus circumstellar continuum emission with wavelength-dependent extinction. A Markov Chain Monte Carlo analysis shows that the most likely model parameters are consistent with a low-temperature, low-gravity photosphere with significant extinction and no more than modest continuum veiling. Its $T_{\mathrm{eff}} \simeq 3260$ K effective temperature is similar to that of older, more evolved pre-main-sequence stars, but its surface gravity log $g \simeq 2.4$ cm s$^{-2}$ is approximately 1 dex lower. This implies that the radius of this protostar is a factor of $\sim 3$ larger than that of $10^6$ yr old T Tauri stars. Its low veiling is consistent with a circumstellar disk having intrinsic near-IR emission that is less than or equal to that of more evolved Class I protostars. Along with the high extinction, this suggests that most of the circumstellar material is in a cold envelope, as expected for a Class 0 protostar. This is the first known detection and analysis of a Class 0 protostar absorption spectrum.

\end{abstract}


\section{Introduction} \label{sec:intro}

The physical stages and observational properties of low-mass protostellar evolution have become clear over the past several decades. The $IRAS$, $ISO$, $Spitzer$, $Herschel$, and other observatories have discovered and characterized the infrared (IR) emissions of thousands of young stars embedded in nearby galactic clouds. Analysis of the IR spectral energy distributions and mm observations of the gaseous environments of these objects have shown that there are now over a hundred known low-mass Class 0 protostars in nearby (within 1 kpc), objects than have yet to accrete the majority of their final masses \citep[see review][]{DSA14}. These are the most deeply embedded accreting objects with central gaseous cores. They have $L_{\rm sub-mm} / L_{\rm bol} \gtrsim 5 \times 10^{-3}$, are surrounded by massive circumstellar envelopes, and are frequently associated with jets or outflows \citep{AWB93}. Much has been learned about the bolometric temperatures \citep{ML93}, luminosities, and lifetimes of the low mass population of these young stellar objects (YSOs) from mostly unresolved far-IR space-based observations \citep[see review by][]{DSA14}. ALMA observations are now revealing high-resolution details of their envelopes and outflows\citep[e.g.,][]{EDL15, AOA17}.

Little is known about the heavily embedded central stars of these objects. \citet{AWB93} noted that VLA 1623, the Class 0 archetype, was undetected at $\lambda < 24$ $\mu$m wavelengths at the time of that study. This was modeled as being due to the extremely high $A_V \gtrsim 1000$ mag extinction of its massive ($\sim 0.6$ $M_\sun$) circumstellar envelope. The $Spitzer$ c2d Legacy survey detected a few dozen Class 0 YSOs \citep{EDJ09} in nearby dark clouds, and many of them have $F_{3.6\mu {\rm m}} \gtrsim 1$ mJy \citep{EES09}. We have found that some of these are detected in sensitive $K$-band ground-based observations including $UKIDSS$ data \citep{LWA07}. Some near-to-mid-IR flux is clearly leaking out through holes in the envelopes that surround these youngest stars.

Some information is known about the central stars of the somewhat more evolved Class I YSOs that are still accreting but have already accumulated the bulk of their final stellar mass. Optical and near-IR spectra have shown that these objects have late-type photospheres with effective temperatures and surface gravities similar to T Tauri stars but sometimes with considerably higher continuum veiling (or IR excess) and faster $v$ sin $i$ rotation velocities \citep[see review by][]{WGD07}. 

What are the natures of the central photospheres of the even more embedded Class 0 protostars? Do they have effective temperatures and surface gravities similar to T Tauri and Class I protostars, or are they lower gravity or otherwise different? Are they more or less veiled than the somewhat more evolved Class I population? Studying these issues would inform how the youngest stars assemble themselves and also probe their currently unresolved warm circumstellar disks. We have started to address these issues with a new study that has detected photospheric features in the spectrum of a Class 0 protostar for the first time. 

The Class 0 source, S68N, lies in the nearby (d=436$\pm$9.2 pc) Serpens
star-forming cloud core \citep{ORT17}. This region was first explored at
near-IR wavelengths leading to the discovery of an embedded cluster of
young stars \citep{SVS76,ECA92,KAS99}. The first millimeter/submillimeter
continuum maps of the 6$^{\prime} \times 5^{\prime}$ Serpens core were made with
the UKT14 bolometer on the JCMT\footnote{The James Clerk Maxwell Telescope is a
15-m diameter submillimeter telescope, which was formerly funded by a partnership
between the U.K., Canada, and the Netherlands. The East Asian Observatory took
over operations in March 2015.}, leading to the identification of nine distinct
1.1-mm continuum sources, some lacking NIR counterparts \citep{CED93,WEC95}. S68N
was not among these. Instead, S68N was first identified due to its strong
molecular line emission \citep{MCM94,HBW96}. Subsequent special high-resolution
processing of the {\it IRAS} survey data enabled construction of source spectral
energy distributions (SEDs) leading to the identification of five distinct Class
0 protostars in the Serpens core, including S68N \citep{HUB96}. The first
detection of submillimeter continuum emission from S68N was from a JCMT 450
$\mu$m map, obtained to ascertain the location of the exciting source of its
compact, CS(2$\rightarrow$1) outflow \citep{WOL98}. \citet{EES09} found that S68N
(also identified there as Ser-emb 8) was detected in all $Spitzer$ IRAC (3.6 --
8.0 $\mu$m) and MIPS 24 and 70 $\mu$m bands with point-source fluxes $\sim$1 mJy
-- 10 Jy. They determined its bolometric temperature to be $T_{\rm bol}$ = 58 K
when all $Spitzer$ fluxes are combined with their 1.1 mm Caltech Submillimeter
Observatory Bolocam photometry.

We describe our new near-IR spectroscopic observations of Serpens S68N and their reduction in Section \ref{sec:data}.  Next we present our analysis of the spectrum in Section \ref{sec:analysis} and discuss its results in Section \ref{sec:discussion}.  Finally, we summarize our conclusions in Section \ref{sec:summary}.

\section{Observations and Data Reduction} \label{sec:data}

We obtained moderate resolution, moderate S/N near-IR $K$--band spectra of the Class 0 protostar Serpens S68N on 2014 June 18 and 19 UT with mostly clear skies and 0\farcs5 -- 0\farcs8 seeing. All observations were made with the Keck~II telescope on Mauna Kea, Hawaii, using its NIRSPEC facility spectrograph \citep{McLeanetal98} in its low resolution, long-slit mode. 

We pointed the telescope to the band-merged $Spitzer$ position of S68N (also identified as Ser-emb 8) given in
Table 3 of \citet{EES09}: $\alpha=$18$^h$29$^m$48.12$^s$,
$\delta=\,+$01$^{\circ}$16$^{\prime}$44.9$^{\prime\prime}$ (J2000). We observed the $K$-band source at that
location and measured the actual pointing from the telescope encoders to be within 1$^{\prime\prime}$ of the
commanded $Spitzer$ position in each coordinate axis. This $K$-band counterpart of S68N appears closest to the
location of SMM 9 (T) in Figure 2 of \citet{H99} and has apparent magnitude $K_s\,=$ 16.1\footnote{K. Haisch,
2018, priv. comm., from KPNO 4-m NEWFIRM photometry acquired in 2013}.

Spectra were acquired with a 0\farcs76 (4-pixel) wide slit, providing spectroscopic resolving power $R \equiv \lambda / \delta \lambda \simeq$ 1000 for PSFs larger than the slit width.  The plate scale was 0\farcs144 pixel$^{-1}$ along the 42$\arcsec$ slit length. The NIRSPEC-7 blocking filter was used to record the $\lambda = 2.06 - 2.46$ $\mu$m wavelength range in each exposure.

The slit was held physically stationary during the exposures and thus rotated on
the sky as the non-equatorially-mounted telescope tracked when observing. Data
were acquired in pairs of exposures of durations of 120 or 150 s each, with the
telescope nodded 10$\arcsec$ or 15$\arcsec$ along the slit between integrations
so that object spectra were acquired in all exposures. The A0 dwarf HD 172792
(HIP 91748) and HD 229700 (HIP 92486) were observed for telluric correction of
the protostar spectra. The telescope was automatically guided with frequent
images from the NIRSPEC internal SCAM IR camera. Spectra of the internal NIRSPEC
continuum lamp were taken for flat fields, and exposures of the Ne and Kr lamps
were used for wavelength calibrations.

All data were reduced with IRAF. First, object and sky frames were differenced
and then divided by normalized flat fields. Next, bad pixels were fixed via
interpolation, and spectra were extracted with the APALL task. Spectra were
wavelength calibrated using low-order fits to lines in the arc lamp exposures,
and spectra at each slit position were co-added at similar airmasses and times.
Instrumental and atmospheric features were removed by dividing
wavelength-calibrated object spectra by spectra of early-type stars observed at
similar airmass at each slit position. Telluric-corrected spectra were produced
by combining the spectra of both slit positions and then multiplying them by the
spectrum of a 10,000 K dwarf stellar model. We summed the multiple individual
spectra to produce a final co-added spectrum, with a total of 248.7 minutes
(14,920 s) of integration time.

\section{Analysis} \label{sec:analysis}

\begin{figure}[ht!]
\plotone{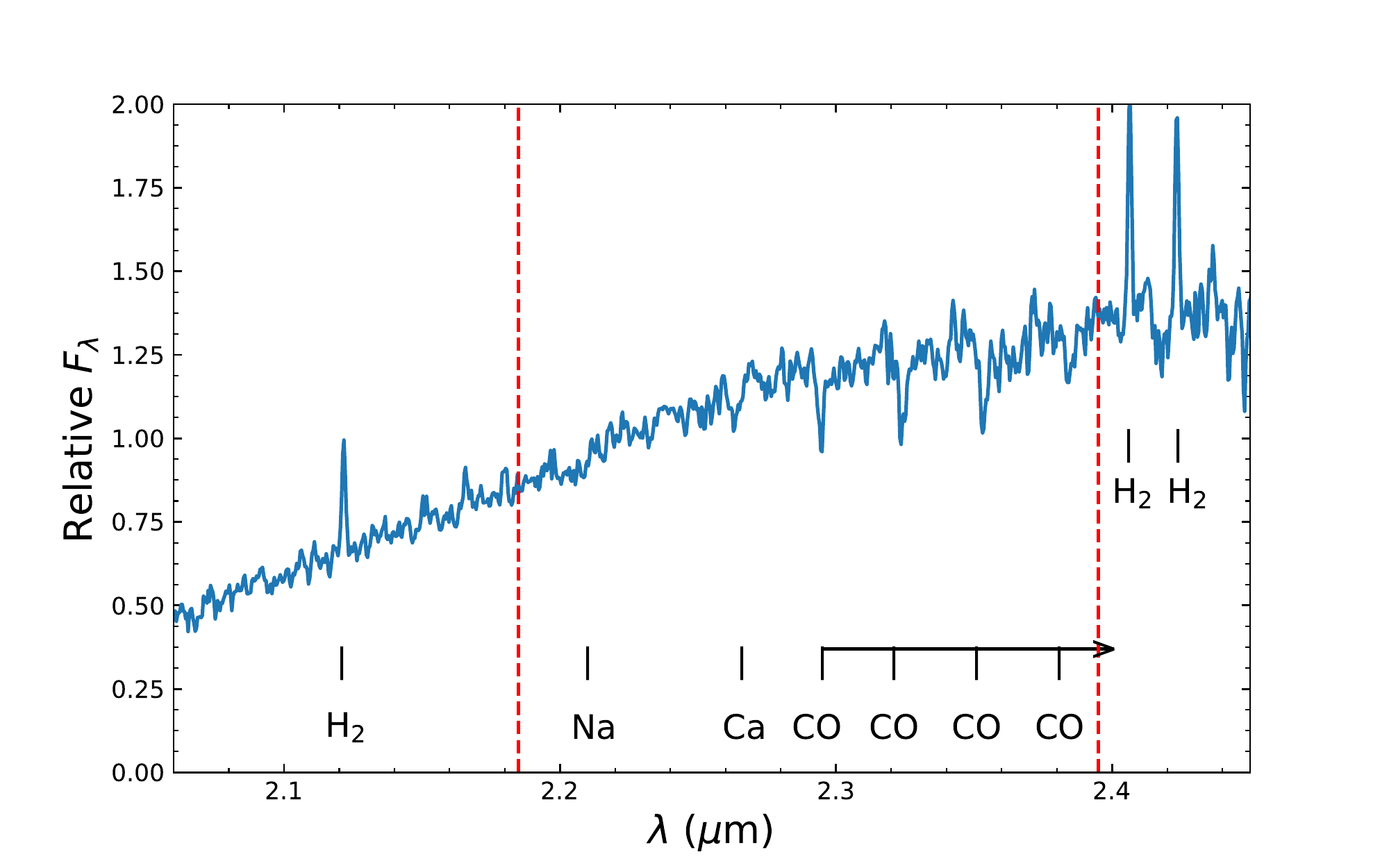}
\caption{$K$--band spectrum of the Class 0 protostar Serpens S68N. Major emission and absorption features are identified and labeled. The region analyzed for properties of the protostellar photosphere and its environment is between the two vertical dashed red lines. The spectrum shown in this figure is available as the Data behind the Figure. \label{fig:spectrum}}
\end{figure}

The final spectrum of Serpens S68N is shown in Figure \ref{fig:spectrum}. This spectrum displays a very red continuum that is punctuated with emission and absorption features. Strong H$_{2}$ $v = 1-0$ $S(1)$, $v = 1-0$ $Q(1)$, and $v = 1-0$ $Q(3)$ emission lines are seen at 2.1218, 2.4066, and 2.4237 $\mu$m respectively. The 2.2477 $\mu$m H$_{2}$ $v = 2-1$ $S(1)$ line is not clearly detected. These and all wavelengths hereafter are for vacuum observations. We measured the equivalent widths and line fluxes of these features and present the results in Table \ref{H2}. The measured H$_2$ FWHM line widths listed in Table \ref{H2} indicate that the emission features have widths of $194 \pm 26$ km s$^{-1}$. This corresponds to a resolving power of $R = 1550$ and is consistent with the quoted $R \simeq 1000$ of this NIRSPEC data given that the seeing was better (less) than or equal to the slit width. \citet{GBW10} found that Class I protostars typically have intrinsic H$_2$ FWHM line widths under 30 km s$^{-1}$, so it is not unreasonable to assume that the S68N H$_2$ lines are strongly unresolved in the spectrum.

\begin{deluxetable}{lrrrr}
\tablecaption{Emission Line Measurements\label{H2}}
\tablewidth{700pt}
\tablehead{
\colhead{H$_{2}$ line} & \colhead{$\lambda$ ($\mu$m)} & 
\colhead{EW\tablenotemark{a} (\AA)} & \colhead{FWHM\tablenotemark{b} (\AA)} & 
\colhead{Line Flux\tablenotemark{c}}\\
} 
\startdata
$v = 1-0$ $S(1)$ & 2.1218 & -8.44 & 15.6   & 55706 \\
$v = 2-1$ $S(1)$ & 2.2477 & $\geq$-0.93\tablenotemark{d} & \nodata &  $\leq$9900\tablenotemark{d} \\
$v = 1-0$ $Q(1)$ & 2.4066 & -6.95 & 13.5   & 96673  \\
$v = 1-0$ $Q(3)$ & 2.4237 & -7.93 & 15.7   & 107660 \\
\enddata
\tablenotetext{a}{Measured equivalent width from Gaussian fitting}
\tablenotetext{b}{Full width half maximum of the best-fit Gaussian. All emission lines were unresolved, with FWHM $<$ $\sim20$\AA\ as expected for the NIRSPEC $R \simeq 1000$ 4 pixel wide slit.}
\tablenotetext{c}{Line flux is the product of the EW and continuum values in instrumental units, not calibrated to physical flux units}
\tablenotetext{d}{The H$_{2}$  $v = 2-1$ $S(1)$ line was not detected. 3$\sigma$ upper limit values are given for EW and Line Flux.} 
\end{deluxetable}

Relatively strong $\delta v = 2$ CO absorption bands are seen over 2.29 -- 2.39 $\mu$m wavelengths, and weak atomic absorption lines of Na I and Ca I may be present in the 2.21 and 2.26 $\mu$m regions. We measured S/N $\simeq$ 30 at a number of continuum wavelengths across the spectrum's $\lambda = 2.06 - 2.46$ $\mu$m wavelength range.

We now describe a likely model for the observed absorption spectrum of Serpens S68N and describe the Markov Chain Monte Carlo (MCMC) analysis we performed to estimate the most likely model parameters.  We created this model to analyze the flux from the central protostar and its modification by its circumstellar environment but not the origin of the H$_2$ emission lines seen in the spectrum. We restricted the wavelengths of the analyzed spectral region to 2.1850 -- 2.3950 $\mu$m (see Figure \ref{fig:spectrum}) in order to ensure that fits to the strongest photospheric absorption features would not be degraded by the relatively high noise in other regions. This region does not include any strong H$_2$ emission lines.

\subsection{Spectral Modeling\label{subsec:modeling}}	

We assume that the nascent photosphere of a Class 0 protostar is surrounded by a circumstellar disk and both are embedded in a large gaseous and dusty envelope within a dark cloud. We model the near-IR protostar spectrum as being the sum of contributions of components due to a stellar photosphere and excess continuum due to warm dust emission in the circumstellar disk. Wavelength-dependent extinction from the envelope and cloud is applied to both emitting components. We do not explicitly include wavelength-dependent scattering in this conceptually simple model; instead its effects are included in the extinction magnitude and its wavelength dependence. This model of the observed flux is described mathematically in equation \ref{eqn:model}:

\begin{equation}\label{eqn:model}
F_{p*,\lambda} = [F_{*,\lambda}(T_{\rm eff}, {\rm log} \: g, {\rm[Fe/H]}) \Omega_{*} + B_\lambda(T_d) \Omega_{\rm d}] \: 10^{-0.4 A_K \left(\frac{\lambda}{\lambda_K}\right )^\alpha}
\end{equation}
 
where $F_{*,\lambda}(T_{\rm eff}, {\rm log} \: g, {\rm[Fe/H]})$ is the flux from the protostellar photosphere of effective temperature $T_{\rm eff}$ surface gravity log $g$, and metallicity [Fe/H]. $\Omega_{*}$ is the solid angle of the protostellar photosphere, $B_\lambda(T_d)$ is the blackbody emission from the circumstellar disk of temperature $T_d$, and $\Omega_{\rm d}$ is the solid angle of the disk emission. $A_K$ is the effective $K$--band extinction with wavelength dependence $(\lambda/\lambda_K)^\alpha$ where $\lambda_K$ is the 2.2 $\mu$m band-center wavelength. Deriving physically meaningful $\Omega_{*}$ and $\Omega_{\rm d}$ parameters would require an absolute spectro-photometric calibration of the observed spectrum and an absolute calibration of the extinction amplitude. Such calibrations would have significant uncertainty, and hereafter we will focus on the $\Omega_d / \Omega_*$ solid angle ratio which is effectively independent of these calibrations.

\subsection{Parameter Estimation Framework\label{subsec:parameters}}	

We employed the Starfish inference framework from \citet{2015ApJ...812..128C} with support for mixture model modifications from \citet{GSHC17}. Starfish provides rapid parameter estimation by swift computation of the likelihood function and efficient MCMC exploration of parameter space. Rapid likelihood computation is enabled by first ingesting a pre-computed model grid and performing a principal component analysis (PCA) of the grid before starting the MCMC process. The results of this PCA are used to quickly compute model values during the information retrieval process, without having to compute complex models each time the likelihood function is evaluated. We had Starfish perform this PCA on the \citet{HWD13} Phoenix stellar models, and then it used the PCA results to compute the photospheric flux $F_{*,\lambda}(T_{\rm eff}, log \: g, [Fe/H])$ in  equation \ref{eqn:model} when evaluating the likelihood function for different model parameters. Starfish uses the emcee \citep{2013PASP..125..306F} package for its MCMC parameter analysis.

We augmented the physical protostar model given in equation (\ref{eqn:model}) with several other parameters that were applied to fit the spectrum of Serpens S68N but were not interpreted to have physical significance in the model (MCMC nuisance parameters). Radial velocity (RV) and spectral width terms were used to allow for slight shifts in the wavelength fit and to allow small departures from the specified instrumental resolving power. Small ($\sim 2$\%) residual differences between the measured absorption spectrum and the equation (\ref{eqn:model}) model were fit with a third-order Chebyshev polynomial, and three Gaussian processes parameters were also applied to allow for fitting a small amount of correlated noise.

We assigned several priors to the model parameter values before performing the MCMC analysis and we provide these in Table \ref{param_table}. The photospheric parameters $T_{\mathrm{eff}}$ and log $g$ were limited to the prior ranges given in that table before the PCA was performed. We restricted [Fe/H] to the solar value in order to minimize the number of model parameters, using [Fe/H] = $0.01 \pm 0.001$. The slight offset from [Fe/H] = 0.0 was made for computational convenience. The extinction (reddening) power law index was limited to $\alpha = -2.0 - -1.7$ to accommodate a range of galactic extinction laws \citep[e.g.,][]{NTH09, MLR05, MW90, RL85}. Any wavelength dependence in extinction in the observed spectrum is likely a combination of true extinction and also scattering off grains (e.g. in cavity walls) into our line of sight. Thus only an {\em effective} extinction power law can be derived from the data, and scattering may be significant if there is not a good fit to the data when applying these prior values. We also limited the range of circumstellar disk temperature, and $A_{k}$ extinction as given in Table \ref{param_table}. 

The RV prior was limited to $\pm 1000$ km s$^{-1}$. The spectral velocity width prior was limited to be within 240 -- 270 km s$^{-1}$, between that of the H$_2$ lines (see the start of this section) and the quoted NIRSPEC resolving power for this mode ($R \simeq 1000$ or 300 km s$^{-1}$). Using a broader prior produced broader posterior distributions of the physical model parameters, perhaps because NIRSPEC and other grating spectrographs have variable resolving power with wavelength  or the impact of noise in the data. We ran Starfish for 20,000 steps, with all parameters converging to limited ranges within the first 10,000 steps.

\subsection{Results\label{subsec:results}}

\begin{figure*}[ht!]
\epsscale{1.2}
\plotone{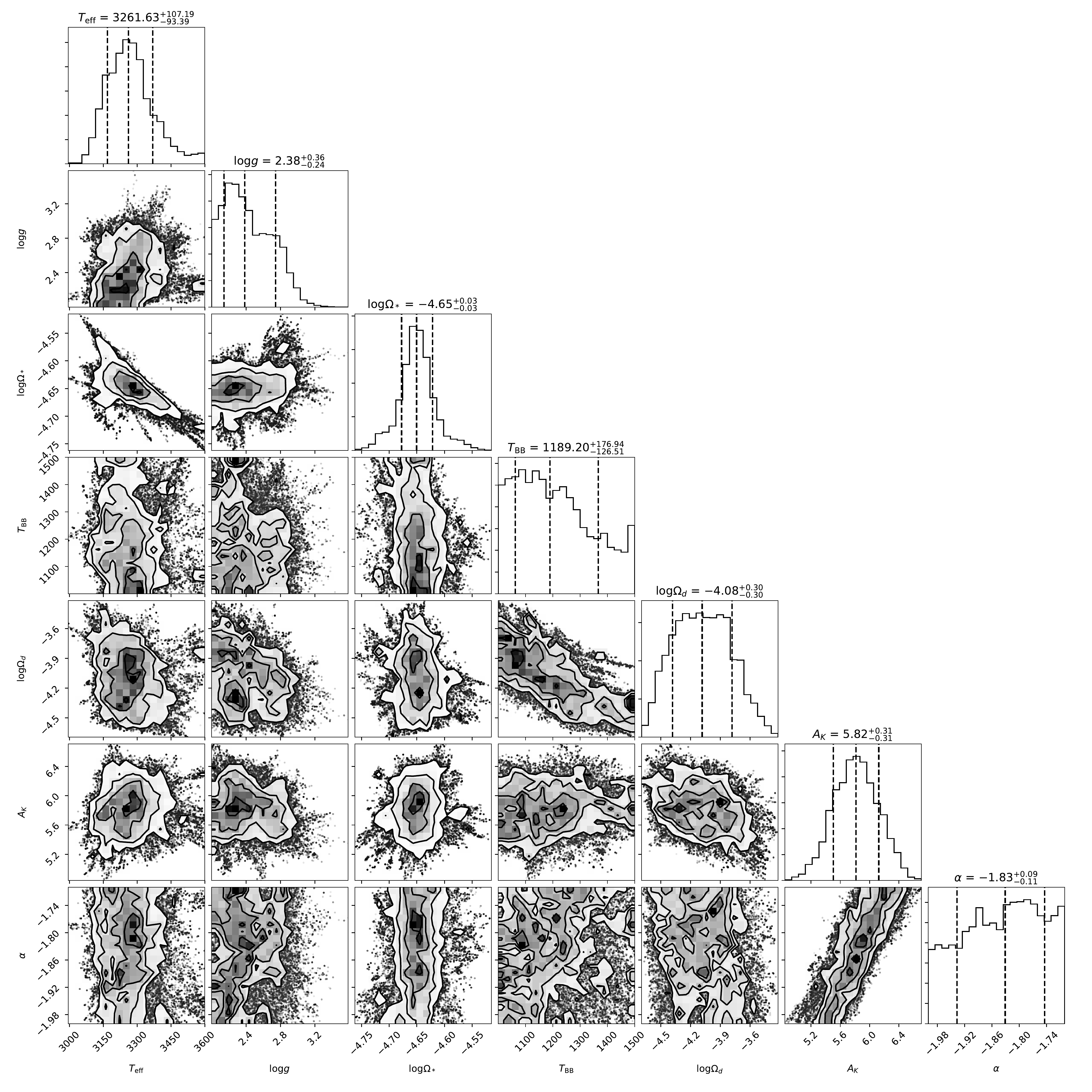}
\caption{Corner plot of S68N posterior distributions and covariances. Values for the last 5000 steps of the 20,000 step Starfish run are shown.  Median values and 68\% distribution points are given above the distribution plot (top) for each parameter shown, and the other plots show correlations between parameters.
\label{fig:S68N_corner}}
\end{figure*}

The MCMC posterior probability distributions and covariances of S68N model
parameters are shown in Figure \ref{fig:S68N_corner}. Table \ref{param_table}
shows the derived median, 16\%, and 84\% (68\% width) confidence values of the
posterior probability distributions for the scientific model (equation
\ref{eqn:model}) parameters of S68N and what priors were applied to the MCMC
analysis. These show that the $T_{\rm eff}$, log $\Omega_*$, log
$\Omega_d$, and $A_K$ parameters are well-constrained with fairly symmetric,
Gaussian-like distributions. Surface gravity is constrained to low values log $g
\lesssim 2.8$, the circumstellar disk blackbody temperature favors values $T_{\rm d}
\lesssim 1400$ K, and the extinction / reddening index is not strongly
constrained. The spectral width parameter had a median value of 261 km s$^{-1}$ but was not well constrained.

\begin{deluxetable}{lrrl}
\tablecaption{Derived Model Parameters and Priors\label{param_table}}
\tablewidth{700pt}
\tablehead{
\colhead{Parameter} & \colhead{S68N\tablenotemark{a}} & 
\colhead{Prior} & \colhead{units}\\
} 
\startdata
$T_{\mathrm{eff}}$ & $3261\substack{+107 \\ -93}$ & 2700 -- 3600 & K\\
log $g$ & $2.38\substack{+0.36 \\ -0.24}$    &  2.0 -- 4.0 & cm s$^{-2}$ \\
$\Omega_d / \Omega_*$ & $3.72\substack{+3.71 \\ -1.87}$  & \nodata & \\
$T_{\rm d}$ & $1189\substack{+177 \\ -127}$      &  1000 -- 1500 & K\\
$A_{k}$ & $5.82\substack{+0.31 \\ -0.31}$    & 0 -- 20 & mag\\ 
$\alpha$ & $-1.83\substack{+0.09 \\ -0.11}$  & -2.0 -- -1.7 & \\
RV\tablenotemark{b}  & $27.0\substack{+12.9 \\ -12.5}$ & -1000 -- +1000 & km s$^{-1}$
\enddata
\tablenotetext{a}{Posterior parameter distribution median $\pm$ 68\% confidence widths.}
\tablenotetext{b}{Uncorrected to LSR or otherwise.}
\end{deluxetable}

\section{Discussion} \label{sec:discussion}

We now discuss the results of the preceding analysis and offer interpretations of the physical nature of the protostellar photosphere its circumstellar material.

\subsection{Photospheric Parameters\label{subsec:photosphere}}	

The most likely parameters of the S68N spectral model (Table \ref{param_table})
indicate that it has a low effective temperature, very low surface gravity and
near-solar metallicity. Its $T_{\mathrm{eff}} \simeq 3261$ K effective
temperature is similar to that of a M3 -- M3.5 II -- V star and is typical of
sub-giant (IV) or dwarf (V) pre-main-sequence (PMS) stars \citep[e.g.,
see][]{PM13}. However, the S68N surface gravity (log $g \simeq 2.38$ cm
s$^{-2}$) is much lower than expected for a typical T Tauri star. For example,
a $10^6$ yr old, 0.20 $M_\odot$ mass, solar composition PMS star
is expected to have $T_{\mathrm{eff}} = 3226$ K and log $g = 3.39$ cm s$^{-2}$
with a radius of $R = 1.49 R_\odot$ \citep{BHA15}. We estimate that the surface
gravity of Serpens S68N is 1.0 dex lower, implying a radius that is $\sqrt{10}$
larger, or $R = 4.7 R_\odot$, if we assume that S68N has the same 0.20
$M_\odot$ mass of the \citet{BHA15} PMS model. This low surface gravity and large radius may be caused by internal heating due to the protostar accreting at a very high rate. \citet{BEV17} showed that accretion bursts can increase the luminosities of classical T Tauri stars by over 0.5 dex compared to a non-accreting star at the same effective temperature. It is not inconceivable that this could be on the order of 1 dex for the more heavily accreting Class 0 protostars. A 1.0 dex increase in luminosity would correspond to a $\sqrt{10}$ increase in radius, consistent with the estimated value of S68N when compared to a PMS model.
 
\subsection{Extinction\label{subsec:extinction}}	

The high extinction of S68N ($A_{k} = 5.82$ mag; see Table \ref{param_table}) corresponds to $A_{v} \simeq 60$ mag depending on the chosen extinction law.  This is consistent with the $A_{v} > 100$ mag extinction expected for Class 0 protostars \citep{AWB93} if its circumstellar envelope has some inhomogeneities which cause non-isotropic extinction. This may be a reasonable assumption since some near-IR flux escapes from the protostar.

We note that the near-IR flux observed from any Class 0 protostar has likely experienced both extinction and scattering by its circumstellar material \citep[e.g., see][]{THC08}. Therefore the extinction value determined from our simple model (equation \ref{eqn:model}) is an {\it effective} value that is impacted by scattering. Figure \ref{fig:S68N_corner} and Table \ref{param_table} show that the MCMC analysis does not provide a very strong constraint on the effective extinction power law $\alpha$. This could be because the competing effects of extinction, scattering, and circumstellar emission are difficult to disentangle over the limited wavelength range of the spectrum.

We investigated whether the H$_2$ emission lines seen in the S68N spectrum (Figure \ref{fig:spectrum}) could be used as an independent measure of extinction to the protostar. The 2.4237 $\mu$m $v = 1-0$ $Q(3)$ line and the 2.1218 $\mu$m $v = 1-0$ $S(1)$ lines have the same upper level, so they have an intrinsic line ratio that is set by only their transition probabilities and is not a function of excitation. This would normally make this line pair an excellent extinction probe; any deviation from the intrinsic ratio \citep[$v = 1-0$ $Q(3)$ / $v = 1-0$ $S(1)$ = 0.7;][]{TKD77, B07} is expected to be due to interstellar extinction.

We measured those H$_2$ lines to have a flux ratio of 1.93 (see Table \ref{H2}), consistent with extinction $A_{k} = 3.51$ mag. This is is 2.31 mag lower than the median $A_{k} = 5.82$ mag estimated by our MCMC analysis (Table \ref{param_table}), equivalent to $\delta$$A_{v} \simeq 23$ mag. The H$_2$ lines straddle the wavelengths used for fitting the protostar extinction and other parameters (2.1850 -- 2.3950 $\mu$m; see Section \ref{sec:analysis} and Figure \ref{fig:spectrum}). It may be possible that extinction is lower to the emission line region if it is located between the protostar photosphere and Earth. In that case, the emission-line extinction estimate would only be a lower limit to the total extinction to the protostar's photosphere. 

However, the H$_2$ emission line extinction estimate is likely to be unreliable for another reason. \citet{CG10} note that there is a strong and narrow telluric absorption line in the Earth's atmosphere at 2.42412 $\mu$m vacuum wavelength. The Earth's orbital velocity causes this line to shift wavelength and overlap with the H$_2$ $v = 1-0$ $Q(3)$ line, reducing its measured flux. This causes the H$_2$ emission line extinction estimate to be underestimated here, but future space-based observations by $JWST$ or other observatories would not have this problem.

\subsection{Circumstellar Emission and Veiling} \label{subsec:veiling}

The circumstellar continuum emission in the spectrum is characterized by its effective dust temperature $T_{\rm d} \simeq 1189$ K and its solid angle ratio, $\Omega_d / \Omega_* \simeq 3.72$ (median values in Table \ref{param_table}). We use these to estimate the $K$-band continuum veiling, $r_k$ = $F_{k, \rm CS}$ / $F_{k,*}$ where $F_{k, \rm CS}$ is the circumstellar continuum flux and $F_{k,*}$ is the stellar flux. We assume that both the star and the circumstellar material emit blackbody emission at their respective temperatures $T_{\mathrm{eff}}$ and $T_{\rm d}$. This approximation simplifies the continuum veiling computation to:

\begin{equation}\label{eqn:veiling}
r_k \simeq \frac{B_\lambda(T_{\rm d})\Omega_{\rm d}}{B_\lambda(T_{\rm eff})\Omega_{*}}.
\end{equation}

We evaluated this veiling equation at $\lambda = 2.2$ $\mu$m to estimate that the continuum veiling in the spectrum is $r_k \simeq 0.10$ This is much lower than the $r_k \sim 0.5 - 3$ typically found for older and less embedded Class I protostars \citep{DGC05, CG10}. Furthermore, we detect $\Omega_d / \Omega_*$ at only the $\sim 2 \sigma$ confidence level (see Table \ref{param_table}), so our measured veiled value of $r_k \simeq 0.10$ is highly uncertain and does not differ from  $r_k = 0$ with high confidence. \citet{WGD07} and \citet{DGC05} found that high resolution visible and near-IR spectra showed that Class I YSOs have effective temperatures and surface gravities similar to T Tauri PMS stars. For a given amount of warm circumstellar disk emission, $r_k$ would be expected to be about an order of magnitude lower for Class 0 protostars if their radii are about $\sqrt{10}$ times larger. Therefore our spectrum is consistent with intrinsic warm circumstellar emission that is similar to or lower than that of Class I YSOs given the low surface gravity of S68N.

We also note that the presence of $\delta v = 2$ CO {\em absorption} bands in the
spectrum is consistent with S68N not having a warm/hot, dense inner disk.
Somewhat more evolved protostars have been found to sometimes have near-IR CO
bands in {\em emission} \citep[e.g.,][]{NCG07, DGC05, CG10}, consistent with hot
(2000 -- 4000 K), dense ($n_{\rm H_2} \sim 10^9$ cm$^{-3}$) circumstellar gas.
Some of these have also been found to have velocity profiles consistent with
rotation in circumstellar disks \citep[e.g.,][]{CTN93, NCG07}. Like some other
Class 0 protostars \citep[e.g., see][]{TCW13}, Serpens S68N may be too young to
have yet developed a massive circumstellar disk, so there could be less warm
material near the central photosphere than in many Class I objects. Regardless,
the high extinction of S68N shows that it is surrounded by a large amount of cold
material, as expected for a Class 0 protostar.

\subsubsection{Nature of H$_2$ line emission} \label{subsec:H2emission}

We now use the measured wavelengths and fluxes of the H$_2$ lines to assess the nature of this emission. We used the measured wavelengths of the H$_2$ lines to compute a mean radial velocity of $-28.1 \pm 3.2$ km s$^{-1}$ (uncorrected to VLSR), so this emission is blue-shifted by $55 \pm 13$ km s$^{-1}$ relative to the best fit photospheric absorption line RV (Table \ref{param_table}). This value is 0.28 times the measured velocity width of the data and is detected at 4 $\sigma$ confidence. We are unsure whether there is any useful information in this result due to its small value and the best fit photospheric absorption line RV may have been influenced by imperfect telluric corrections or other factors.

Finally, we assess the excitation of the circumstellar H$_2$ emission by evaluating
the strengths of its observed lines. \citet{GD95} showed that the ratio of
H$_{2}$ $v = 1-0$ $S(1)$ to $v = 2-1$ $S(1)$ is relatively sensitive to
excitation mechanisms. They computed the ratios of these 2 lines to be 1.9 for UV
excitation, 7.7 for shocked gas at $T=2000$ K, and 16.7 for X-ray excitation of
low ionization H$_{2}$. Using the H$_2$ line fluxes in Table \ref{H2} and the
median $\alpha = -1.83$ extinction power law from Table \ref{param_table}, we
find that S68N's reddening-corrected $v = 1-0$ $S(1)$ to $v = 2-1$ $S(1)$ line
flux ratio is greater than or equal to 6.33. This value is consistent with shock
or X-ray but not UV excitation of H$_{2}$ using the \citet{GD95} values above. It
is difficult to distinguish between shocked and X-ray excited H$_{2}$ emission
using only a single near-IR line ratio because collisions will thermalize the
excitation levels of H$_{2}$ in a sufficiently dense gas \citep[e.g., see the
discussion in][]{GBW10}. However, the measured ratio is sufficient to determine
that the H$_2$ emission lines in S68N are not produced by pure UV excitation
\citep{BD87, GD95}.

\section{Summary and Conclusions} \label{sec:summary}

We have detected weak photospheric absorption lines and H$_2$ emission lines in the near-IR, $K$-band spectrum of the Class 0 protostar Serpens S68N. We performed a MCMC analysis on the absorption spectrum and determined that the protostar's photosphere has an effective temperature similar to more evolved PMS stars, but its surface gravity is on the order of 1 dex lower. This low gravity implies a large stellar radius, which may be consistent with recent and possibly episodic heavy accretion. The absorption spectrum is consistent with a large amount of effective extinction (combined extinction and scattering), as expected for a Class 0 object with a large amount of circumstellar material. Its near-IR H$_2$ emission is consistent with excitation in shocks or by X-rays but not by UV radiation. The near-IR continuum veiling is considerably lower than measured for more evolved Class I protostars and is not detected at high confidence. This is understandable given the much larger radius of the central photosphere, and it may also indicate that the there is little dust or gass mass in the inner, warm regions of the circumstellar disk. This plus the large amount of extinction suggests that most of the circumstellar material is in a cold envelope, as expected for a Class 0 protostar.

The moderate data quality of this faint object limited the precision of our analysis. We expect that $JWST$ will produce superior quality data (higher S/N and resolving power) that will be able to constrain the properties of this and other Class 0 protostars with considerably higher precision.

\acknowledgments
We thank Isabelle Baraffe, Peter Martin, Ewine van Dishoeck, Brunella Nisini
and Ian Czekala for helpful discussions and encouragement which inspired and improved this work. We also thank an anonymous referee for a thoughtful and helpful review. TPG acknowledges support for this work from NASA WBS 411672. This work was also supported by a NASA Keck PI Data Award, administered by the NASA Exoplanet Science Institute. Data presented herein were obtained at the W. M. Keck Observatory from telescope time allocated to the National Aeronautics and Space Administration through the agency's scientific partnership with the California Institute of Technology and the University of California. The Observatory was made possible by the generous financial support of the W. M. Keck Foundation.

The authors wish to recognize and acknowledge the very significant cultural role and reverence that the summit of Mauna Kea has always had within the indigenous Hawaiian community. We are most fortunate to have the opportunity to conduct observations from this mountain.

\vspace{5mm}
\facility{Keck:II (NIRSPEC)}

\software{  corner \citep{2016JOSS.2016...24F},
			emcee \citep{2013PASP..125..306F},
			IRAF \citep{1993ASPC...52..173T},
			Starfish\citep{2015ApJ...812..128C}
			}

\end{document}